\documentclass[aps,prl,reprint,floatfix,twocolumn,amsmath,amssymb,superscriptaddress,longbibliography]{revtex4-1} 
\usepackage{graphicx}
\usepackage{color}
\usepackage{dcolumn}
\usepackage{soul}
\usepackage{currfile}
\usepackage[normalem]{ulem}

\newcommand{\Ang}{\AA$^{-1}$}

\newcommand{\XQE}{\ensuremath{\chi^{\prime\prime}(Q,E)}}
\newcommand{\vesi}{BaCu$_3$V$_2$O$_8$(OH)$_2$}
\newcommand{\vesid}{BaCu$_3$V$_2$O$_8$(OD)$_2$}

\newcommand{\volb}{Cu$_3$V$_2$O$_7$(OH)$_2 \cdot 2$H$_{2}$O}
\newcommand{\kap}{ZnCu$_3$(OH)$_{6}$Cl$_{2}$}
\newcommand{\hay}{MgCu$_3$(OH)$_{6}$Cl$_{2}$}

\begin{document}

\title{Vesignieite: a $S = \frac{1}{2}$ kagome antiferromagnet with dominant third-neighbor exchange}
\author{D.~Boldrin}
   \email[Corresponding author: ]{d.boldrin@imperial.ac.uk}
   \altaffiliation[Present address: ]{Department of Physics, Blackett Laboratory, Imperial College London, London, SW7 2AZ, United Kingdom}
   \affiliation{Department of Chemistry, University College London, 20 Gordon Street, London, WC1H 0AJ, United Kingdom}
\author{B.~F\aa k}
   \affiliation{Institut Laue-Langevin, 71 avenue des Martyrs, CS 20156, 38042 Grenoble Cedex 9, France}
\author{E. Can\'evet}
   \affiliation{Institut Laue-Langevin, 71 avenue des Martyrs, CS 20156, 38042 Grenoble Cedex 9, France}
   \affiliation{Laboratoire de Physique des Solides, CNRS, Univ.\ Paris-Sud, Universit\'e Paris-Saclay, 91405 Orsay Cedex, France}
\author{J.~Ollivier}
   \affiliation{Institut Laue-Langevin, 71 avenue des Martyrs, CS 20156, 38042 Grenoble Cedex 9, France}
\author{H. C. Walker}
   \affiliation{STFC Rutherford Appleton Lab, ISIS Facility, Harwell Science and Innovation Campus, Didcot, OX11 0QX, United Kingdom}
\author{P. Manuel}
   \affiliation{STFC Rutherford Appleton Lab, ISIS Facility, Harwell Science and Innovation Campus, Didcot, OX11 0QX, United Kingdom}
\author{D. Khalyavin}
   \affiliation{STFC Rutherford Appleton Lab, ISIS Facility, Harwell Science and Innovation Campus, Didcot, OX11 0QX, United Kingdom}
\author{A.~S.~Wills}
   \affiliation{Department of Chemistry, University College London, 20 Gordon Street, London, WC1H 0AJ, United Kingdom}

\date{\today \quad \currfilename}

\begin{abstract}
The spin-$\frac{1}{2}$ kagome antiferromagnet is an archetypal frustrated system predicted to host a variety of exotic magnetic states. We show using neutron scattering measurements that deuterated vesignieite \vesid, a fully stoichiometric $S=1/2$ kagome magnet with $<$1\% lattice distortion, orders magnetically at $T_{\mathrm{N}}=9$~K 
into a multi-{\bf k} coplanar variant of the predicted triple-{\bf k} octahedral structure. We find this structure is stabilized by a dominant antiferromagnetic 3$^{\mathrm{rd}}$-neighbor exchange $J_3$ with minor 1$^{\mathrm{st}}$- or 2$^{\mathrm{nd}}$--neighbour exchange. 
The spin-wave spectrum is well described by a $J_3$-only model including 
a tiny symmetric exchange anisotropy.
\end{abstract}
\maketitle

Geometrically frustrated magnets have the potential to host a multitude of exotic ground states, such as the elusive quantum spin liquid (QSL) states \cite{Anderson1987,Balents2010,Savary2017}. Having been studied theoretically for several decades, experimental progress has been hampered by a lack of model materials. However, recent experiments on new systems have revealed novel excitations and emergent phenomena beyond those originally predicted of both fundamental and technological interest \cite{Banerjee2017,Fak2017,Klanjsek2017,Paddison2017}. The $S=1/2$ kagome lattice is an archetypal frustrated system due to the combination of quantum spins and the low connectivity of its corner-sharing triangular lattice. Initial interest and understanding of this system focused on the nearest-neighbor (NN) Heisenberg Hamiltonian,
as found experimentally in the well studied Herbertsmithite \cite{Han2012,Norman2016}, spurred by the potential to realize various flavors of QSLs \cite{Yan2011,Iqbal2014}. 
However, as most experimental realizations of the kagome antiferromagnet show
deviations from a simple isotropic NN model, recent research has been extended to include exchange anisotropy and
further neighbor exchanges \cite{Cepas2008,He2015,Messio2011}. 
This has turned out to be equally interesting, 
with theory predicting novel types of magnetic order, such as octahedral and cuboc structures for classical spins \cite{Messio2011} and chiral or U(1) spin liquids for quantum spins \cite{Bieri2015,He2014}. 
The vast majority of these theoretical predictions await experimental verification.

Model systems through which the effects of further neighbor interactions can be explored in $S = \frac{1}{2}$ kagome magnets are rare. Kapellasite \kap\ \cite{Fak2012} and haydeeite \hay\ \cite{Boldrin2015hay} are both members of the paratacamite mineral system shared by herbertsmithite,  but have ferromagnetic NN exchange. In the former, antiferromagnetic further neighbor interactions drive the system into a chiral QSL state with fluctuations born from a cuboc2 phase \cite{Fak2012,Bieri2015}, 
while in the latter antiferromagnetic further neighbor interactions are again present but fail to drive the system away from the ferromagnetic ground state  \cite{Boldrin2015hay}. 
In volborthite, \volb, the distortion of the kagome lattice leads to inequivalent NN interactions and different local CuO$_{6}$ environments, suppressing long-range order to very low temperatures \cite{Nilsen2011,Yoshida2012volbo} 
and leading to a thermal Hall effect in  the spin liquid state \cite{Watanabe2016}.

The mineral vesignieite \vesi\ \cite{Guillemin1955} is a copper vanadate mineral similar to volborthite with a fully stoichiometric yet distorted $S = \frac{1}{2}$ kagome lattice, although this distortion is significantly less in the former, $<$1\% in terms of bond distances. Given this comparison and considering the strong antiferromagnetic interactions inferred from the Weiss temperature $\theta_{\mathrm{W}} \sim -80$~K and initial findings consistent with a quantum spin liquid state \cite{Okamoto2009,Zhang2010,Colman2011,Quilliam2011,Okamoto2011}, it is surprising that vesignieite is now known to order magnetically at $T_\mathrm{N}=9$~K \cite{Yoshida2012,Yoshida2013,Boldrin2016,Ishikawa2017}. However, a more complete comparison of the system with both theory and related experimental systems requires determination of the magnetic structure and exchange interactions.

\begin{figure}
\includegraphics[width=0.95\columnwidth]{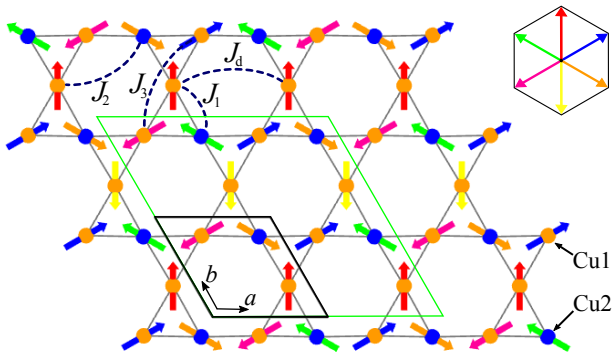}
\caption{ 
The kagome lattice of vesignieite showing the two Cu$^{2+}$ sites and the first four exchange interactions. 
The arrows on the Cu sites show the hexagonal magnetic structure with ${\bf k}=\{1/2,0,0\}$. In this structure, each moment, represented by the arrow color, has Fourier components associated with only a single \textbf{k}-vector. The black line is the crystallographic unit cell and the green line is the magnetic unit cell. The inset illustrates how the spins point towards the vertices of a hexagon.}
\label{FigLattice}
\end{figure}

In this Letter, we use neutron scattering to determine the magnetic propagation vector {\bf k} and the magnetic excitation spectrum of vesignieite. 
We find that the material adopts an unconventional non-collinear multi-${\bf k}$ structure with ${\bf k} =\{1/2,0,0\}$, stabilized by a dominant antiferromagnetic third-neighbor $J_3$ interaction. We term this a hexagonal structure based on the nomenclature of the regular magnetic orders on the kagome lattice \cite{Messio2011}, see Fig.\ \ref{FigLattice}, and find it is closely related to the theoretically predicted non-coplanar octahedral structure.

Vesignieite consists of nearly perfect Cu$^{2+}$ kagome layers that are separated by VO$_4$ tetrahedra and Ba$^{2+}$ ions. 
Several  crystallographic structural models have been proposed  \cite{Yoshida2012,Ishikawa2017,Okamoto2009,Colman2011,Yoshida2012,Yoshida2013,Zhesheng1991,Boldrin2016}. 
The appearance of small satellite Bragg peaks in X-ray powder diffraction measurements in deuterated powder samples suggest the trigonal $P3_121$ space group (No.\ 152), where the tiny scalene distortion of the triangles correspond to less than 1\% difference in the three Cu \--- Cu bond lengths \cite{Boldrin2016}.
The lowering of the kagome symmetry is due to a cooperative Jahn-Teller distortion and leads to the formation of two distinct Cu$^{2+}$ sites as shown in Fig.\ \ref{FigLattice}, where the (orange) Cu2 site is orbitally disordered whilst the (blue) Cu1 site is orbitally ordered \cite{Boldrin2016}. 

We used the fully deuterated  \vesid\  single-phase powder samples of Ref.\ \cite{Boldrin2016}. 
Magnetic susceptibility measurements show a negative value of  $\theta_{\mathrm{W}}=-75$~K  \cite{Boldrin2016}, in good agreement with other vesignieite samples 
\cite{Okamoto2009,Colman2011,Yoshida2012,Boldrin2016,Ishikawa2017},
 and which suggests predominant antiferromagnetic interactions. 
However, 
attempts to estimate the strength and anisotropy of the exchange interactions based on nearest-neighbor exchange $J_1$ only \cite{Okamoto2009,Yoshida2012,Zorko2013} are bound to fail, as vesignieite has a dominant $J_3$ exchange as we will show further below. 
Also, high-temperature series expansions for the kagome lattice \cite{Bernu2015} are not known for this particular case. 

\begin{figure}
\includegraphics[width=0.75\columnwidth]{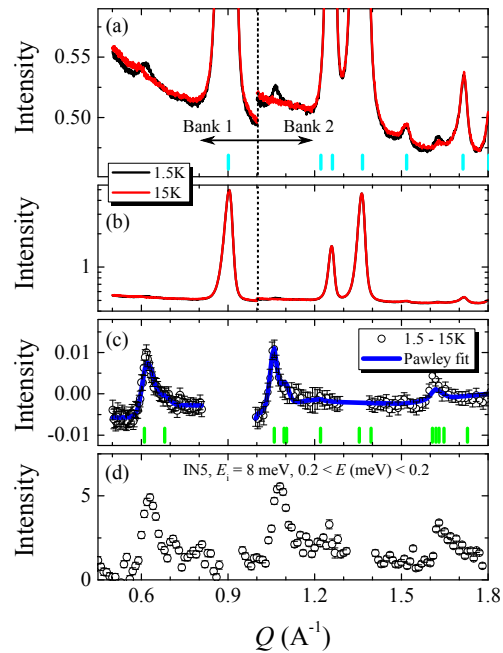}
\caption{
Neutron diffraction data collected on banks 1 and 2 of the WISH diffractometer showing (a) weak magnetic Bragg peaks at $T = 1.5$~K (black) compared to $T=9$~K (red) and (b) the full intensity scale.
Blue tick marks indicate the  nuclear reflections from the $P3_121$ crystal structure. The vertical dashed lines separate the data from banks 1 and 2. 
(c) Magnetic diffraction at $T = 1.5$~K after subtraction of the $T=15$~K data fitted using a Pawley refinement (blue line). 
Green tick marks indicate the reflections from the ${\bf k}=(1/2,0,0)$ magnetic structure.
(d) Elastic line from INS data of the IN5 spectrometer at $T = 1.5$~K after subtraction of $T=30$\,K data showing the same three magnetic Bragg peaks. 
}
\label{FigMagBragg}
\end{figure}

Our sample shows a clear antiferromagnetic phase transition at $T_\mathrm{N}=9$~K, 
as in other high-quality powder samples \cite{Yoshida2012,Yoshida2013,Boldrin2016}. 
A small zero-field-cooled/field-cooled splitting in the magnetic susceptibility \cite{Yoshida2012,Yoshida2013,Boldrin2016,Ishikawa2017} 
and a tiny hysteresis loop in the magnetization curve \cite{Yoshida2012,Boldrin2016,Ishikawa2017} suggest the formation of a weak ferromagnetic component with a magnitude of 1.6\% of the staggered moment, 
possibly related to a minute spin canting within the kagome planes. 

Neutron diffraction data collected on WISH at ISIS show weak additional Bragg peaks indicative of a magnetically ordered phase at $T=1.5$~K when compared with the paramagnetic phase at $T=25$~K, see Fig.~\ref{FigMagBragg}(a--c), with an intensity corresponding to about 0.2~$\mu_B$. 
Inelastic neutron scattering (INS) measurements on the time-of-flight spectrometer IN5 at the ILL also show such additional peaks at elastic energy transfers below $T_\mathrm{N}$, see Fig.~\ref{FigMagBragg}(d). 
A Pawley fit to the neutron diffraction data of Fig.~\ref{FigMagBragg}(c) shows that these peaks, 
observed at $Q$ values of 0.64, 1.08, and 1.65~\Ang, can be indexed by a propagation vector of ${\bf k}=(1/2,0,0)$, which corresponds to the M point of the Brillouin zone of the kagome lattice. 

The M point is of particular interest, as it characterizes the three regular triple-{\bf k} magnetic structures on the kagome lattice, namely the octahedral, cuboc1, and cuboc2 states  \cite{Messio2011}. 
These states can be distinguished by their extinction rules \cite{Messio2011}, 
and the observation of the three above mentioned Bragg peaks suggest
that the magnetic structure of vesignieite is related to the octahedral state, 
which is favored by a large antiferromagnetic $J_3>0$ interaction and characterized by 90$^{\circ}$ rotations between neighboring spins \cite{Messio2011}. 
The octahedral state is energetically degenerate with a number of coplanar states, such as the hexagonal (see Fig.\ \ref{FigLattice}), where neighboring spins are at 60$^{\circ}$ or 120$^{\circ}$ with respect to each other.

\begin{figure}
\includegraphics[width=0.95\columnwidth]{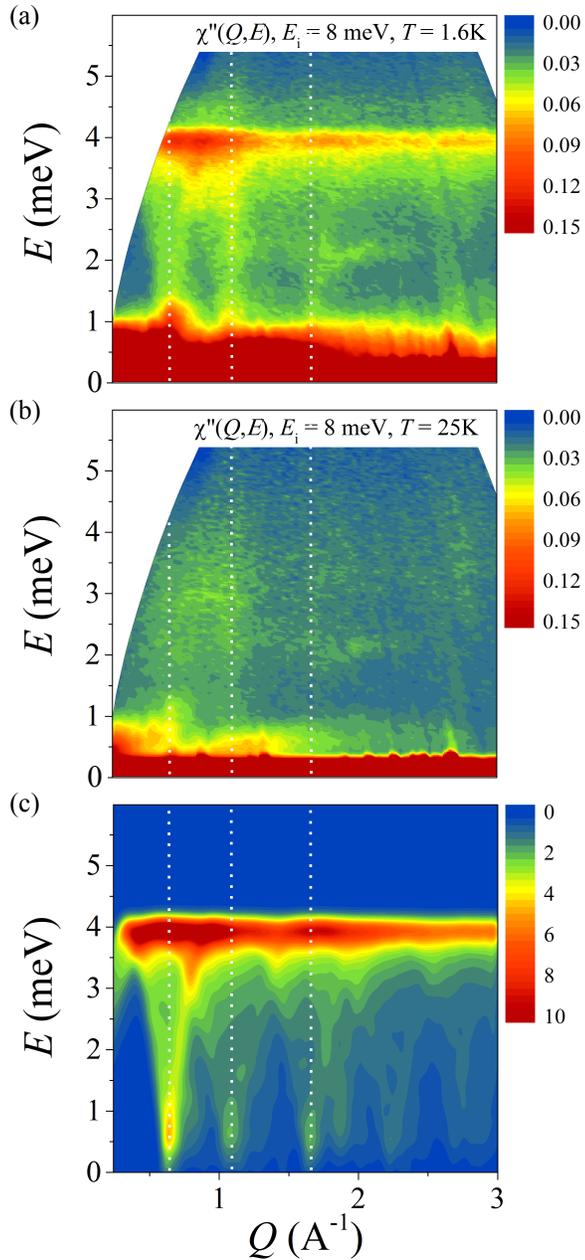}
\caption{
Dynamic susceptibility \XQE\ measured on IN5 with $E_i=8.0$~meV at a temperature of (a) $T=1.6$~K and (b) $T=25$~K. (c) Spin-wave dispersion calculated within linear spin-wave theory with $J_3=1.9$ and $J_{\mathrm{d}} = -0.1$\,meV for an isotropic Hamiltonian. Dashed white lines are guides to the eye at $Q = 0.64$, 1.08, and 1.64~\AA$^{-1}$.
}
\label{FigINS}
\end{figure}

Inelastic neutron scattering measurements were performed on fully deuterated powder samples of weight between 2 and 7~g for temperatures $1.4<T<50$~K and incoming neutron energies $1.9<E_i<88$~meV using the MERLIN (ISIS) and the IN4 and IN5 (ILL) spectrometers. 
In the ordered phase, spin-wave excitations are observed with gapless modes emerging from the magnetic Bragg peak positions discussed above and a strong flat band centered at an energy of 4~meV, see Fig.\ \ref{FigINS}(a). 
These well-defined excitations have completely disappeared at temperatures well above $T_\mathrm{N}$, see Fig.\ \ref{FigINS}(b).

The data were modeled using semi-classical spin-wave theory \cite{Toth2015} 
assuming the hexagonal magnetic structure of Fig.\ \ref{FigLattice} and a spin Hamiltonian
\begin{displaymath}
\mathcal{H} =
\sum_{i,j}  J_{ij} [ S_i^x S_j^x  + S_i^y S_j^y + (1+\delta) S_i^z S_j^z]
+  {\bf D}_{ij} \cdot ({\bf S}_i \times {\bf S}_j) .
\label{EqHam}
\end{displaymath}
The flat mode at $E=4$~meV fixes the main exchange integrals (here given in units of meV) 
to  $J_3 - J_d \approx 2.0$ 
whereas the width and the lack of dispersion of the flat mode impose $|J_d|<0.2$, $|J_1|<0.1$, and $|J_2|<<0.1$. 
The best fit using an isotropic version of the above Hamiltonian 
($\delta=0$ and ${\bf D}_{ij}=0$) was obtained with $J_3=1.9$ and $J_d=-0.1$; the powder-averaged calculation of this model is shown in Fig.\ \ref{FigINS}(c). Identical results were obtained using the octahedral magnetic structure.

A $J_{3}$ dominant model in vesignieite is unexpected given the structural similarities to other $S = \frac{1}{2}$ kagome materials where NN exchange is dominant \cite{Nilsen2014,Han2012,Boldrin2015hay,Fak2012}. Jahn-Teller induced axial compression of the CuO$_{6}$ octahedra in vesignieite 
is due to a singly occupied $d_{\mathrm{x}^{2}-\mathrm{y}^{2}}$ orbital, meaning that both $\mu_{3}$-O(H) and $\mu_{2}$-O sites may be active in NN superexchange. 
The corresponding Cu\---O\---Cu angles range between 101 and $105^{\circ}$ for the former and 78 and $89^{\circ}$ for the latter \cite{Burns1996,Boldrin2015,Boldrin2016}. These values 
lie either side of the expected switchover from antiferro- to ferromagnetic superexchange according to the Goodenough-Kanamori-Anderson rules for copper oxides \cite{Mizuno1998}. It is therefore conceivable that these competing paths cancel in vesignieite, leading to the negligible $J_{1}$ and dominant $J_{3}$ exchange energies we observe.

We will now discuss the hexagonal structure in more detail. It can be formed by projecting the octahedral structure onto 2 dimensions along the cubic [111] direction. 
It minimizes the $J_3$-only antiferromagnetic Heisenberg Hamiltonian on the kagome lattice 
(along with a full continuum of magnetic orders including the octahedral and related coplanar and colinear orders), and it remains lowest in energy for $|J_i|$ up to about $|J_i|\lesssim J_3/2$ 
for $i=1,2,d$.
It is a rather particular triple-{\bf k} structure:
the Fourier components of a given arm of the star of ${\bf k}=\{1/2,0,0\}$ 
[i.e., ${\bf k}_1=(1/2,0,0)$, ${\bf k}_2=(0,1/2,0)$, or ${\bf k}_3=(-1/2,-1/2,0)$] 
are non-zero for only one site. 
In other words, each site has its own {\bf k} vector and its own Fourier component, 
and thus there is no coupling between the ordered moments on different sites. 
This is also clear from Fig.~\ref{FigLattice}: 
an antiferromagnetic $J_3$-only model 
results in three decoupled magnetic lattices. 
Introduction of a $J_d$ interaction, the strongest ``perturbation'' in our case,
does not couple these lattices either, 
but induces some frustration for an antiferromagnetic $J_d$. 
In order to couple the three lattices, either $J_1$ or $J_2$ (the latter negligible in our case) is needed, both of which are frustrating. 

In such a ``decoupled'' multi-{\bf k} structure, 
disordering one of the sites will have no effect on the magnetic order 
nor on the magnetic excitations. 
This is exactly what happens in vesignieite, whereby
the distortion leading to the rare $P3_{1}21$ structural symmetry is a consequence of orbital disorder on the Cu2 site \cite{Boldrin2016}. The partial orbital ordering occurs because the Cu$^{2+}$ $d_{x^{2}-y^{2}}$ orbitals cannot be arranged symmetrically around the kagome triangles, 
unlike the superexchange mediating $d_{z^{2}}$ orbitals in Herbertsmithite \cite{Boldrin2016}. 
Globally, the trigonal symmetry is retained via the $3_1$ screw axis, 
which results in three kagome layers within each unit cell with adjacent layers rotated by 120$^\circ$. Such a scenario is consistent with a multi-\textbf{k} magnetic structure, where the screw axis can now be understood to arise from a ``deselection'' of different arms of the star of {\bf k} in each layer, \emph{i.e.} the disordered site and its associated \textbf{k}-vector changes in adjacent layers.
A similar scenario is found in a partially ordered $S=2$ kagome antiferromagnet  \cite{Ling2017}, but the deduced 
single-\textbf{k} magnetic structure is not consistent with either the trigonal symmetry or the dominant $J_{3}$ exchange of vesignieite, 
and it is not known which exchanges stabilize this structure. 

Further evidence of a multi-\textbf{k} structure in vesignieite is given by the observation of a small ferromagnetic component below $T_{\mathrm{N}}$ \cite{Boldrin2016,Ishikawa2017}. This component, which must be a secondary order parameter, is only consistent with a multi-\textbf{k} ground state. This general conclusion, which does not depend on the direction of the ferromagnetic component, comes from the requirement of time-reversal and translational invariance of the relevant free-energy coupling term. The invariance can only be achieved if the term involves a product of the magnetic order parameter components related to all three arms of the star of {\bf k}.
Furthermore, magnetic susceptibility measurements on a single crystal sample indicate that the spins lie in the kagome plane \cite{Ishikawa2017}. 
The hexagonal structure is consistent with this finding and, 
unlike energetically degenerate collinear structures, also satisfies the trigonal crystal symmetry. Unfortunately, the powder-averaged spin-wave spectra of the hexagonal structure cannot be distinguished within the experimental resolution from that of the non-coplanar octahedral structure.

We will now discuss the energetic stabilization of a coplanar, rather than non-coplanar, structure. The former is favored by a DM interaction on the strongest bond $J_3$ with the DM vector perpendicular to the kagome planes. 
This is allowed in the $P3_121$ space group, but such a DM interaction has no visible effect on the spin-wave spectrum, although it may explain the very weak ferromagnetic in-plane component inferred from magnetic susceptibility measurements \cite{Ishikawa2017}.
A DM interaction on the $J_1$ bond would also have little effect, as $J_1$ is very weak in vesignieite. 
The tiny lattice distortion implies that the $J_3$ exchange along the Cu1--Cu2 chains could be slightly different from that along the  Cu1--Cu1 chains, see Fig.\ \ref{FigLattice}. 
However, this leads only to minute changes in the spin-wave spectrum, 
even for differences as large as 20\% between the two $J_3$ values.  

\begin{figure}
\includegraphics[width=\columnwidth]{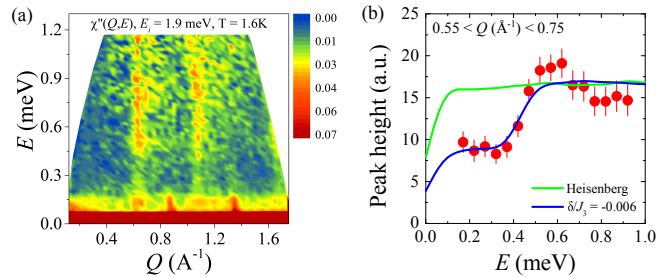} 
\caption{(a) Dynamic susceptibility \XQE\ measured on IN5 with $E_i=1.9$~meV at $T=1.6$~K. (b) Energy scan of \XQE\ at constant $Q=0.65 \pm 0.1$~\Ang, plotted with spin-wave calculations for a coplanar structure with $J_3=2$~meV
for isotropic exchange (green curve) 
and  symmetric exchange anisotropy $\delta/J_3=-0.006$ (blue curve).
 }
\label{FigEscan}
\end{figure}

Instead we find the drive for coplanarity is a symmetric exchange anisotropy on $J_3$, 
with the anisoptropy term $\delta<0$ in the above Hamiltonian. Such an anisotropy lifts one of the two ``acoustic'' branches to create a finite energy gap for that mode (the other mode remains gapless), and this introduces a strong spectral change in the powder average. We find evidence of such a gap in the low-energy dynamic susceptibility \XQE, see Fig.\ \ref{FigEscan}(a). Constant $Q$-cuts of \XQE\ at the acoustic branch $Q=0.65 \pm 0.1$~\Ang\ reveal a significant reduction in intensity below $E \sim 0.5$\,meV, which is in excellent agreement with calculations performed with symmetric exchange anisotropy, see Fig.\ \ref{FigEscan}(b).

In conclusion, our neutron scattering data show that the mineral vesignieite orders magnetically at $T_\mathrm{N}=9$~K in a hexagonal multi-{\bf k} structure closely related to the octahedral structure predicted for the kagome lattice with dominant antiferromagnetic third-neighbor interactions, $J_3>0$. 
The measured spin-wave spectrum is also consistent with such a model.
A particularity of this multi-{\bf k} structure, given that first- and second neighbor interactions are weak compared to third-neighbor interactions, is that it consists of three essentially decoupled lattices, and is therefore insensitive to the orbital disorder of the Cu2 site of vesigniete.
For a Heisenberg Hamiltonian, a dominating $J_3$ would stabilize a continuum of energetically equivalent magnetic structures, including the theoretically predicted  regular octahedral state as well as non-regular coplanar or collinear states. In vesignieite, a weak symmetric exchange anisotropy of $\delta/J_3=-0.006$ 
stabilizes the coplanar hexagonal magnetic structure from this manifold. To satisfy both the anisotropy and the trigonal crystal symmetry, neighboring spins are rotated by 60$^{\circ}$ or 120$^{\circ}$ and point towards the vertices of a hexagon. 
We note that the hexagonal structure can be thought of as three interlocked skewed square lattices, which extends the relevance of this system beyond its kagome geometry.
The consequences these findings have on the recently predicted topological magnetic excitations in vesignieite remain to be seen \cite{Owerre2017}.

\begin{acknowledgments}
This work was supported in part by the 
French Agence Nationale de la Recherche Grant No.\ ANR-12-BS04-0021. 
DB would like to thank Lesley Cohen for continued financial support.
Experiments at the ISIS neutron and muon source was supported by a beam-time allocation from the Science and Technology Facilities Council, UK. 
We thank Sandor Toth, Mike Zhitomirsky, Laura Messio, and Bernard Bernu for helpful discussions.
\end{acknowledgments}

%

\end{document}